\documentclass[aps,prc,twocolumn,superscriptaddress,showpacs]{revtex4}

\usepackage{graphicx}
\usepackage{dcolumn}
\usepackage{bm}
\usepackage{ulem}
\usepackage{color}

\newcommand{\al}{$\alpha$}
\newcommand{\g}{$\gamma$}
\newcommand{\raa}{($\alpha$,$\alpha$)}
\newcommand{\raX}{($\alpha$,$X$)}

\newcommand{\rag}{($\alpha$,$\gamma$)}
\newcommand{\ran}{($\alpha$,n)}
\newcommand{\rann}{($\alpha$,2n)}

\newcommand{\stot}{$\sigma_{\rm{reac}}$}
\newcommand{\sred}{$\sigma_{\rm{red}}$}
\newcommand{\pras}{$^{141}$Pr}
\newcommand{\pmiii}{$^{143}$Pm}
\newcommand{\pmiv}{$^{144}$Pm}
\newcommand{\pmv}{$^{145}$Pm}

\newcommand{\ppro}{p-process}
\newcommand{\sfact}{S-factor}

\begin{document}

\title{
  Total reaction cross sections from $^{141}$Pr($\alpha$,$\alpha$)$^{141}$Pr
  elastic scattering and $\alpha$-induced reaction cross sections at low
  energies
}

\author{Peter Mohr}
\email{WidmaierMohr@t-online.de}
\affiliation{
Diakonie-Klinikum, D-74523 Schw\"abisch Hall, Germany}
\affiliation{
Institute of Nuclear Research (ATOMKI), H-4001 Debrecen, Hungary}

\date{\today}

\begin{abstract}
Elastic scattering data for $^{141}$Pr($\alpha$,$\alpha$)$^{141}$Pr have been
analyzed to derive a new energy-dependent local potential for the
$^{141}$Pr-$\alpha$ system. This potential is used successfully to predict the
cross section of the $^{141}$Pr($\alpha$,n)$^{144}$Pm reaction at low energies
where new experimental data have become available very recently. Contrary to
various global potentials, this new potential is able to reproduce
simultaneously elastic scattering data around and above the Coulomb barrier
and reaction data below the Coulomb barrier for the $^{141}$Pr-$\alpha$
system. Reasons for the partial failure of the global potentials are
explained by intrinsic properties of the scattering matrix and their variation 
with energy. The new local potential may become the basis for the construction
of a new global $\alpha$-nucleus potential.
\end{abstract}


\maketitle

\section{Introduction}
\label{sec:intro}
The total reaction cross section \stot\ is related to the complex scattering
matrix $S_L = \eta_L \, \exp{(2i\delta_L)}$ by the well-known relation
\begin{equation}
\sigma_{\rm{reac}} = \sum_L \sigma_L 
   = \frac{\pi}{k^2} \sum_L (2L+1) \, (1 - \eta_L^2)
\label{eq:stot}
\end{equation}
where $k = \sqrt{2 \mu E_{\rm{c.m.}}}/\hbar$ is the wave number,
$E_{\rm{c.m.}}$ is the energy in the center-of-mass (c.m.)\ system, and
$\eta_L$ and $\delta_L$ are the real reflexion coefficients and scattering
phase shifts. $\sigma_L$ is the contribution of the $L$-th partial wave to the
total reaction cross section \stot .

Usually, experimental elastic scattering angular distributions
are analyzed using a complex optical potential. At first view, it seems to be
a simple task firstly to determine \stot\ from the analysis of the elastic
scattering angular distribution and secondly to distribute this cross section
\stot\ among the open channels (e.g.\ using the statistical model) to predict
cross sections of \al -induced reactions. However, in practice several
problems appear. There is no unambiguous way to determine reflexion
coeffcients $\eta_L$, phase shifts $\delta_L$, or the optical potential from a
measured elastic scattering angular distribution, and, in addition, in most 
cases angular distributions are measured at relatively high energies whereas
reaction cross sections should also be known at low energies below the Coulomb
barrier (this holds in particular for reaction cross sections relevant for
nuclear astrophysics). Thus, typically an ambiguous optical potential has to be
extrapolated down to 
low energies; as a consequence, considerable uncertainties have been noticed
for the prediction of \al -induced reaction cross sections at low energies, in
particular for 
\rag\ capture reactions for targets with masses above $A \approx 100$
\cite{Som98,Gyu06,Ozk07,Cat08,Yal09,Gyu10,Kis11}. 

Very recently, Sauerwein {\it{et al.}}\ \cite{Sau11} have studied the \pras
\ran \pmiv\ reaction at energies between 10 and 15\,MeV, i.e.\ below the
Coulomb barrier. It is shown in \cite{Sau11} that the calculated \pras \ran
\pmiv\ cross section depends almost exclusively on the \al\ transmission and
is thus well-suited to test global \al -nucleus optical potentials. It is
found that the new experimental data cannot be reproduced by any of the widely
used global \al -nucleus potentials which are the energy-independent
4-parameter McFadden/Satchler (MCF) potential \cite{McF66}, the latest version
of the Avrigeanu (AVR) potential with many partly energy-dependent parameters
\cite{Avr10}, and the energy-independent 6-parameter potential by Fr\"ohlich
and Rauscher (FRR) which is optimized for low-energy reaction cross sections
\cite{Rau03}. However, an excellent description of the new \pras \ran
\pmiv\ data is achieved in \cite{Sau11} using an energy-dependent modification
of the MCF potential where a new energy dependence of the depth of the imaginary
Woods-Saxon volume potential was introduced to reproduce the new reaction
data; the potential by Sauerwein {\it{et al.}}\ is referred to as
Sauerwein/Rauscher (short ``SAR'') in the following. 

Contrary to the study in \cite{Sau11} which is restricted to the analysis of
the \pras \ran \pmiv\ reaction in a narrow energy window, the present study
considers \pras \raa \pras\ elastic scattering in a wide energy range from 19
to 45\,MeV. From the fits to the elastic scattering angular distributions a
new energy-dependent potential is derived, and the total reaction cross
section \stot\ is calculated from this potential. \stot\ is then compared to
all in the EXFOR data base \cite{EXFOR} available \al -induced reaction data
on \pras . The aim of the present study is thus a consistent description of all
available elastic scattering and reaction data over a broad energy range.

Elastic \pras \raa \pras\ scattering data are available in literature at
45\,MeV \cite{Baer71}. However, these data cover only a limited angular range,
and they have to be digitized from Fig.~3 in \cite{Baer71}. The latter leads
to uncertainties which are difficult to estimate. Four \pras \raa
\pras\ angular distributions at $E_{\rm{lab}}$ = 19.0, 23.97, 32.0, and
37.7\,MeV have been measured by \cite{Gua94}. The experiment has been
performed at the XTU Tandem of the INFN Laboratori Nazionali di
Legnaro.
Unfortunately, these data have never been published; a partial analysis of the
data was already shown in an earlier publication \cite{Atz96}. These data are
available in numerical form (including uncertainties) and are thus much
better suited for the 
determination of an optical potential by a fitting procedure. However, as the
original data give statistical errors only (with tiny uncertainties at forward
angles), a 5\,\% uncertainty has been added quadratically to account for
unknown systematic errors of the data. Note that it is difficult to achieve
much smaller systematic uncertainties for \al -scattering experiments (see
e.g.\ \cite{Kis09,Kis11a}). These data will be made available to the community
via EXFOR \cite{EXFOR}.

Besides the \pras \ran \pmiv\ data of \cite{Sau11}, the reactions \pras \ran
\pmiv\ and \pras \rann \pmiii\ have been measured in \cite{Afz05} using a
stacked-foil activation technique. The data cover an energy range from about
15 to 45\,MeV. Although this technique leads to considerable uncertainties at
low energies, these data provide further insight into the energy dependence of
reaction cross sections and the distribution of the total reaction cross
section \stot\ among different open channels. Unfortunately, data for \pras
\rag \pmv\ are not 
available in \cite{EXFOR}, and because of its lower reaction cross section and
unfavorable half-life and decay branches of $^{145}$Pm it was not possible to
measure the \pras \rag \pmv\ cross section simultaneously with the \pras \ran
\pmiv\ cross section in the recent high-sensitivity activation experiment of
\cite{Sau11}.

\section{Analysis of elastic scattering and the total reaction cross section
  \stot }
\label{sec:scat}
\subsection{Elastic scattering}
\label{sec:el}
The analysis of the \pras \raa \pras\ angular distributions follows closely
the procedure outlined in earlier work \cite{Kis09,Kis11a}. The total
potential is composed of the nuclear potential with a real and an imaginary
part and the real Coulomb potential. The different potentials are discussed in
the following.

The real part of the nuclear potential is derived from the folding model; the
folding potential is modified by a strength parameter $\lambda \approx 1.2 -
1.4$ and a width parameter $w \approx 1.0$. (Large deviations from $w \approx
1$ would indicate a failure of the folding model.) To avoid discrete
uncertainties from the so-called ``family problem'' real potentials with
volume integrals of about $J_R \approx 320 - 350$\,MeV\,fm$^3$ have been
selected \cite{Atz96}. (Note that the negative signs of volume integrals $J_R$
and $J_I$ are -- as usual -- omitted in the discussion.)

The imaginary part of the nuclear potential is taken in the usual Woods-Saxon
parametrization. It is well-known that scattering data at low energies are
best described using an imaginary potential of Woods-Saxon surface type (see
e.g.\ \cite{Gal05}). This has been confirmed recently in a microscopic
calculation of the \al -nucleus potential \cite{Guo11}. The same behavior is
found in the present study where the angular distributions at $E_{\rm{lab}}$ =
19.0, 23.97, and 32.0\,MeV can be very well reproduced with a pure surface
imaginary potential. An additional volume Woods-Saxon potential in the
imaginary part does not improve the description of the angular distributions
at low energies. However, for an excellent description of the 37.7 and
45.0\,MeV data a combination of volume and surface Woods-Saxon potentials for
the imaginary part is required; at these energies the volume part is even
dominating the imaginary potential.

The Coulomb potential is taken in the usual parametrization of a
homogenously charged sphere with a Coulomb radius $R_C$ identical to the
root-mean-square (rms) radius of the folding potential without width
modification ($w=1$). This avoids uncertainties from the otherwise somewhat
arbitrary choice of the Coulomb radius $R_C$ (often taken as $R_C = 1.2$, 1.3,
or 1.4\,fm $\times A_T^{1/3}$) which are non-negligible at least at
very low energies \cite{Rau11}. 

\begin{figure}[htb]
\includegraphics[bbllx=25,bblly=10,bburx=568,bbury=785,width=\columnwidth,clip=]{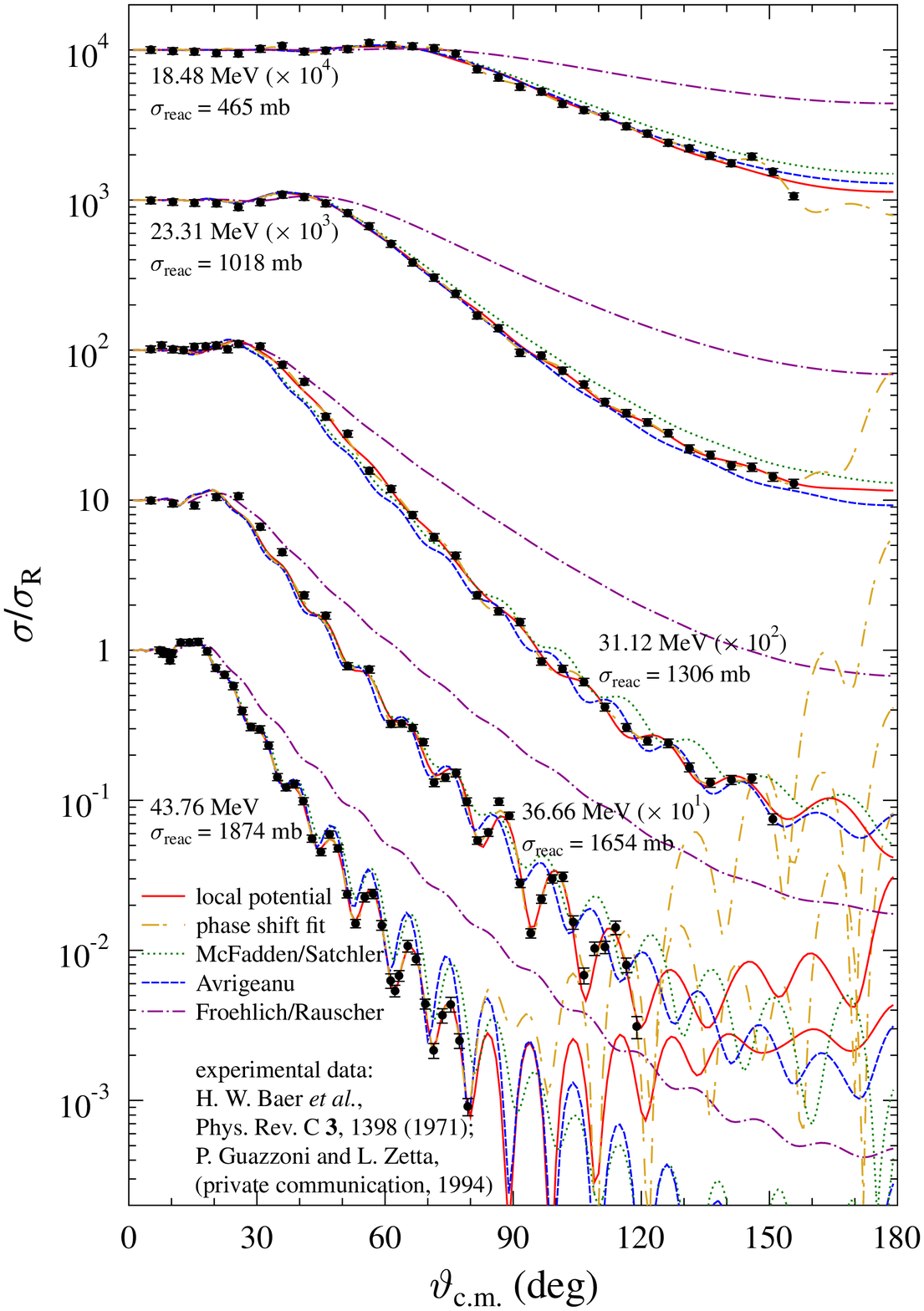}
\caption{
\label{fig:scat}
(Color online)
Rutherford normalized elastic scattering cross sections of the \pras \raa
\pras\ reaction versus the angle in center-of-mass frame. The lines are
calculated from a local potential fit which is adjusted to the scattering data
(full red line), from a phase shift analysis (dash-dotted lightbrown line)
\cite{Chi96}, and from different global \al -nucleus potentials 
\cite{McF66,Avr10,Rau03}. The experimental data have been taken from
\cite{Gua94,Baer71}. The given energies are $E_{\rm{c.m.}}$ in the
center-of-mass system.
}
\end{figure}

The resulting parameters of the potential and the total
reaction cross section \stot\ are listed in Table \ref{tab:pot}. The fits are
compared to the experimental angular distributions in Fig.~\ref{fig:scat}.

\begin{table*}[tbh]
\caption{\label{tab:pot}
Parameters of the optical potential and the total reaction
cross section \stot\ derived from \pras \raa \pras\ angular
distributions.
}
\begin{center}
\begin{tabular}{cccccccccrcccccr@{$\pm$}lc}
\multicolumn{1}{c}{$E_{\rm{lab}}$} 
& \multicolumn{1}{c}{$E_{\rm{c.m.}}$}
& \multicolumn{1}{c}{$\lambda$}
& \multicolumn{1}{c}{$w$}
& \multicolumn{1}{c}{$J_R$}
& \multicolumn{1}{c}{$r_{R,{\rm{rms}}}$}
& \multicolumn{1}{c}{$W_V$}
& \multicolumn{1}{c}{$R_V$}
& \multicolumn{1}{c}{$a_V$}
& \multicolumn{1}{c}{$W_S$}
& \multicolumn{1}{c}{$R_S$}
& \multicolumn{1}{c}{$a_S$}
& \multicolumn{1}{c}{$J_I$}
& \multicolumn{1}{c}{$r_{I,{\rm{rms}}}$}
& \multicolumn{1}{c}{$\chi^2/F$}
& \multicolumn{2}{c}{\stot\ \footnote{from the local potential fit using
    Eq.~(\ref{eq:stot}); uncertainties estimated from the model-independent
    phase shift analysis}} 
& \multicolumn{1}{c}{Ref.} \\
\multicolumn{1}{c}{(MeV)} 
& \multicolumn{1}{c}{(MeV)}
& \multicolumn{1}{c}{(--)}
& \multicolumn{1}{c}{(--)}
& \multicolumn{1}{c}{(MeV\,fm$^3$)}
& \multicolumn{1}{c}{(fm)}
& \multicolumn{1}{c}{(MeV)}
& \multicolumn{1}{c}{(fm)}
& \multicolumn{1}{c}{(fm)}
& \multicolumn{1}{c}{(MeV)}
& \multicolumn{1}{c}{(fm)}
& \multicolumn{1}{c}{(fm)}
& \multicolumn{1}{c}{(MeV\,fm$^3$)}
& \multicolumn{1}{c}{(fm)}
& \multicolumn{1}{c}{(--)}
& \multicolumn{2}{c}{(mb)} 
& \multicolumn{1}{c}{Exp.} \\
\hline
19.0 & 18.48 & 1.354 & 0.994 & 345.3 & 5.511 
& \multicolumn{3}{c}{--} & 17.1 & 1.358 & 0.581 & 45.1 & 7.451 
& 2.3 & 465 & 31 & \cite{Gua94} \\ 
23.97 & 23.31 & 1.270 & 1.010 & 338.6 & 5.604 
& \multicolumn{3}{c}{--} & 27.5 & 1.410 & 0.508 & 68.2 & 7.623 
& 0.8 & 1018 & 31 & \cite{Gua94} \\ 
32.0 & 31.12 & 1.452 & 0.958 & 326.7 & 5.314 
& \multicolumn{3}{c}{--} & 36.1 & 1.290 & 0.500 & 73.9 & 7.012 
& 2.4 & 1306 & 39 & \cite{Gua94} \\ 
37.7 & 36.66 & 1.328 & 0.992 & 329.3 & 5.504 
& -17.8 & 1.297 & 0.157 & 3.9 & 1.586 & 0.457 & 51.7 & 6.075 
& 5.9 & 1654 & 50 & \cite{Gua94} \\
45.0 & 43.76 & 1.306 & 1.012 & 340.0 & 5.614 
& -18.4 & 1.303 & 0.107 & 6.1 & 1.567 & 0.436 & 58.5 & 6.253 
& 0.4 & 1874 & 56 & \cite{Baer71} \\
\multicolumn{2}{c}{low energies} 
               & 1.354 & 0.994 & 345.3 & 5.511 
& \multicolumn{3}{c}{--} & Eq.~(\ref{eq:fermi}) 
& 1.353 & 0.530 & Eq.~(\ref{eq:fermi}) & 7.361 
& -- & \multicolumn{2}{c}{--} & -- \\
\multicolumn{2}{c}{\al\ decay} 
               & 1.151 & 1.000\footnote{fixed}
& 308.8 & 5.545 
& \multicolumn{3}{c}{--} & \multicolumn{3}{c}{--} & --  & -- 
& -- & \multicolumn{2}{c}{--} & -- \\
\hline
\end{tabular}
\end{center}
\end{table*}

In addition to the local potential analysis, a model-independent
phase shift analysis (PSA) has
been performed using the technique of \cite{Chi96}. These phase shift fits
show the tendency to relatively high and oscillating cross sections at
backward angles where no experimental data are available. Nevertheless, the
derived total reaction cross sections \stot\ are close to the results of the
local potential fit. From the variation of \stot\ with 
the fitting parameters (in particular the maximum fitted angular momentum
$L_{\rm{max}}$) and from the comparison with the local potential fit, the
uncertainty of \stot\ can be estimated to be smaller than 3\,\% in all cases
except the lowest energy where I estimate an uncertainty of about 7\,\%. This
is also consistent with a recent analysis of total reaction cross sections in
\cite{Mohr10}. 

As discussed above, the extraction of the total reaction cross section
\stot\ requires theoretical considerations and is thus somewhat
model-dependent. Nevertheless, because of the small sensitivity of \stot\ to
the chosen model, the total reaction cross section \stot\ can be considered as
a quasi-experimental quantity. This holds in particular for cases where the
elastic scattering angular distributions cover the full angular
range. However, as a word of caution, it should be kept in mind that
discrepancies have been noticed between \stot\ determined from elastic
scattering angular distributions and from \al -transmission experiments
\cite{Abe94,Auce94}. These discrepancies have not been fully understood up to
now \cite{Ait95}. 

Furthermore, Fig.~\ref{fig:scat} shows the results of several global \al
-nucleus optical potentials \cite{McF66,Avr10,Rau03}. It is obvious that the
global potentials cannot achieve the same quality as the local potential fit
or the phase shift fit. This is not a surprise because the parameters of the
global potentials are not 
readjusted to the experimental angular distributions. Nevertheless the AVR
potential reproduces the angular 
distributions very well. The calculations using the MCF potential are also
close to the experimental angular distributions; however, there is a
systematic overestimation of the scattering cross sections at backward
angles at low energies. A strong overestimation at backward angles is found
for the FRR potential at all energies; this corresponds to a significant
underestimation of the total reaction cross sections \stot\ of the FRR
potential. It has to be pointed out here that the FRR potential was never
intended to reproduce elastic scattering data above the Coulomb barrier.

The present study does not show results of the potentials of Kumar {\it et
  al.}\ \cite{Kum06} because this potential has been optimized for a wide
energy range above the new \pras \ran \pmiv\ data; in addition, it has been
shown in \cite{Kis09} that this potential cannot reproduce $^{89}$Y\raa
$^{89}$Y elastic scattering data at low energies.
Unfortunately, the latest versions \cite{Dem09,Dem07} of the global 
potential by Demetriou {\it et al.}~\cite{Dem02} are only published in
conference proceedings and cannot be used without the authors of
\cite{Dem09,Dem07}; I do not intend to show results from the early and
perhaps out-dated potentials in \cite{Dem02}. Not yet included are also the
predictions from a new regional potential which was derived from recent
scattering data of the nuclear astrophysics group at Notre Dame university; a
publication is in preparation \cite{Pal12}.

\subsection{Reduced cross sections \sred }
\label{sec:sred}
\begin{figure}[htb]
\includegraphics[width=\columnwidth,clip=]{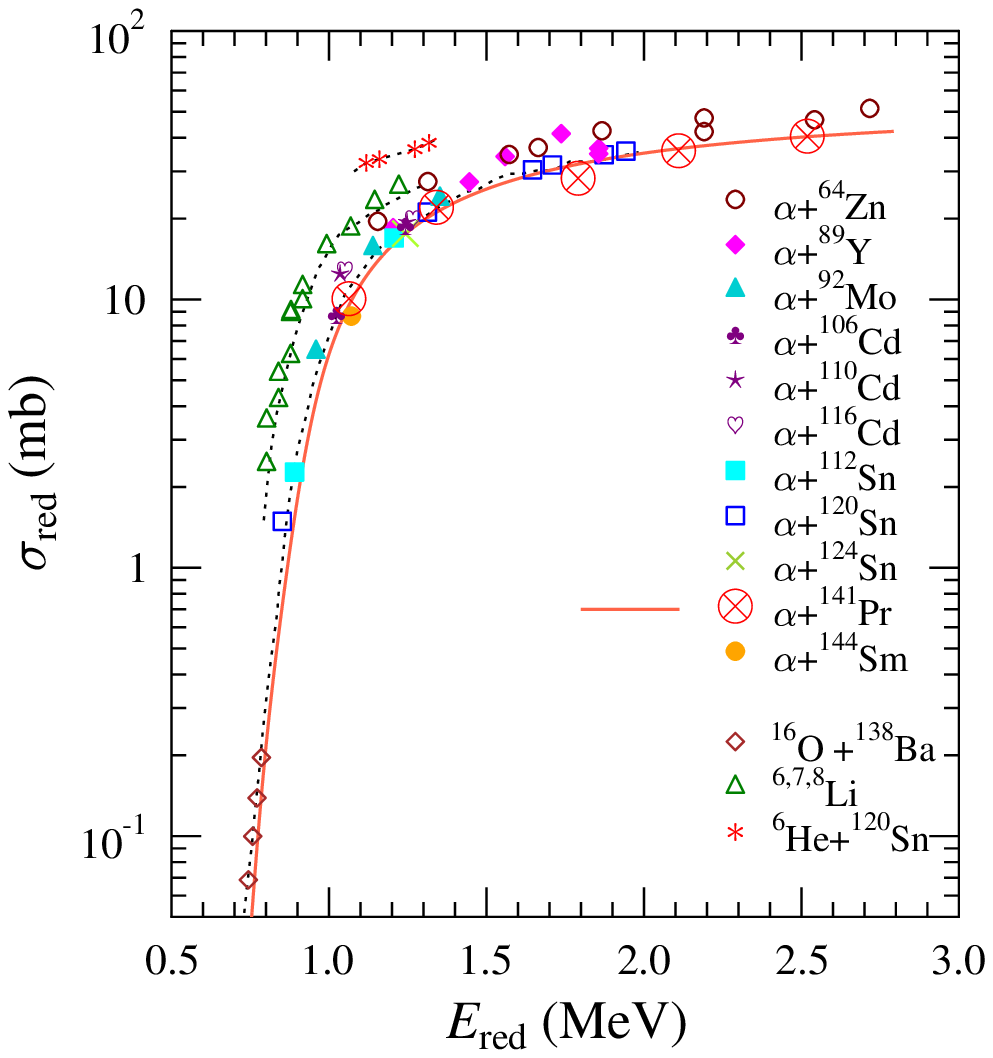}
\caption{
\label{fig:sig_red}
(Color online)
Reduced reaction cross sections $\sigma_{\rm{red}}$ versus reduced energy
$E_{\rm{red}}$ for tightly bound \al -particles and $^{16}$O, weakly bound
$^{6,7,8}$Li projectiles, and exotic $^{6}$He. (Update of Fig.~4 from
\cite{Far10} with additional data from
\cite{Mohr10,Mohr10a,Kis09,Kis11a,Mohr12}). The error bars of the new data
for \pras\ (huge red symbols) are omitted because they are smaller than 
the point size. The dotted lines are to guide the eye. The full red line is
calculated from the energy-dependent local potential for \pras -\al\ (see
Sect.~\ref{sec:energy}). 
}
\end{figure}

For a comparison of total reaction cross sections for different nuclei at
various energies, it has been suggested to present the data as so-called
``reduced cross sections'' versus ``reduced energies'' as defined by
\begin{eqnarray}
E_{\rm{red}} & = & \frac{\bigl(A_P^{1/3}+A_T^{1/3}\bigr) E_{\rm{c.m.}}}{Z_P Z_T} \\
\sigma_{\rm{red}} & = & \frac{\sigma_{\rm{reac}}}{\bigl(A_P^{1/3}+A_T^{1/3}\bigr)^2}
\label{eq:red}
\end{eqnarray}
The reduced energy $E_{\rm{red}}$ takes into account the different heights of
the Coulomb barrier in the systems under consideration, whereas the reduced
reaction cross section \sred\ scales the measured total reaction cross section
\stot\ according to the geometrical size of the projectile-plus-target system.
A smooth behavior for all \sred\ of \al -induced reactions is found, including
the new data for \pras -\al . The
obtained values for \sred\ are smaller for tightly bound projectiles (\al ,
$^{16}$O) compared to weakly bound projectiles ($^{6,7,8}$Li) and halo
projectiles ($^6$He). The data are shown in Fig.~\ref{fig:sig_red} which is an
update of similar figures in \cite{Far10,Mohr10}. Up to now, no complete
theoretical analysis has been presented for the reduced cross sections
\sred\ in Fig.~\ref{fig:sig_red}, and for better readability the data points
have been connected by lines ``to guide the eye'' (dotted lines in
Fig.~\ref{fig:sig_red}; taken from \cite{Far10}). In addition to these lines,
I show here the result from the local potential for \pras \raa \pras\ with its
energy dependence as discussed in the next section
(Sect.~\ref{sec:energy}). This calculation reproduces practically all data
points for \al -induced reactions and also data for $^{16}$O-$^{138}$Ba which
is another combination of doubly-magic projectile and semi-magic target
nucleus. 

The smooth behavior of all reduced cross sections \sred\ in
Fig.~\ref{fig:sig_red} encourages to search for a global potential which is
able to reproduce \sred\ and thus the energy dependence ot the total reaction
cross sections \stot\ for \al -induced reactions. The present study was
restricted to \pras\ but an extension to a wider target range is planned for
the near future. The following procedure is applied to derive an
energy-dependent \al -\pras\ potential from the local potential fits (see
also Table \ref{tab:pot}).

\subsection{Energy dependence of the potential}
\label{sec:energy}
Here I discuss the extraction of a local energy-dependent potential for \pras
-\al . An estimate of the uncertainties for the resulting total reaction cross
section \stot\ will be given later in Sect.~\ref{sec:sensitivity}.

The energy dependence of the real part is weak. The volume integrals $J_R$
in Table \ref{tab:pot}
increase slightly with decreasing energy; however, at even lower energies the
opposite behavior is suggested from dispersion relations. Thus, the real part
of the potential is simply taken from the lowest angular distribution at
19\,MeV where the width parameter $w$ is close to 1.0 and close to the average
of the other energies. Note that the parameters of the real part vary only
weakly with energy. This holds also for the potential derived from the
analysis of the weak \al -decay branch of \pmv\ (see Sec.~\ref{sec:decay}).

The situation for the imaginary part is more difficult because the parameters
vary with energy, and different parametrizations had to be used for the five
angular distributions under study. For the extrapolation to low energies I use
a surface Woods-Saxon potential (parameters ``low energies'' in
Table~\ref{tab:pot}) with the average geometry from the angular distributions
at 19.0, 23.97, and 32.0\,MeV, i.e.\ $R_S = 1.353$\,fm and $a_S =
0.530$\,fm. Only the three lowest energies are considered here because these
three angular distributions could be described with the same parametrization
(surface Woods-Saxon) of the 
imaginary part. Note that the radius parameters $R_S$ and the diffuseness
parameters $a_S$ do not vary significantly with energy, i.e.\ the shape of the
imaginary potential is well defined from the experimental angular
distributions. (A significant energy dependence of the
shape of the imaginary part is found only for halo-like projectiles like
$^6$He, see e.g.\ \cite{Mohr10a}.)

The strength $W_S$ of the imaginary
potential is derived by fitting the imaginary volume integral $J_I$ of the
three lowest energies using a Fermi-type function (similar to \cite{Som98} and
\cite{Sau11})
\begin{equation}
J_I = \frac{J_{I,0}}{1+\exp{[(E_0-E_{\rm{c.m.}}/a_E)]}}
\label{eq:fermi}
\end{equation}
with the parameters $J_{I,0} = 74.16$\,MeV\,fm$^3$, $E_0 = 17.41$\,MeV, and
$a_E = 2.42$\,MeV; a similar formula like Eq.~(\ref{eq:fermi}) holds for the
depth $W_S$ of the surface imaginary potential with $W_{S,0} =
31.1$\,MeV. A similar Fermi-type function was also used in \cite{Sau11} for the
depth of the imaginary volume Woods-Saxon potential. It is interesting to note
that the parameters $E_0$ and $a_E$ in Eq.~(\ref{eq:fermi}) are close to the
values obtained from the adjustment to the new \pras \ran \pmiv\ data in
\cite{Sau11}. 

The widely used Brown-Rho parametrization of the imaginary part \cite{Bro81}
is given by
\begin{equation}
J_I = J_{I,0} \, \frac{(E_{\rm{c.m.}} - E_0)^2}{(E_{\rm{c.m.}} - E_0)^2 + \Delta^2}
\label{eq:BR}
\end{equation}
for energies $E_{\rm{c.m.}} > E_0$ above the opening of the first non-elastic
channel at energy $E_0$ and $J_I = 0$ below $E_0$. $J_{I,0}$ is again the
saturation value, and the parameter $\Delta$ describes the slope of $J_I$ 
from zero to its saturation value $J_{I,0}$. $E_0$ is given by the excitation
energy of the lowest excited state in \pras\ which can be populated by inelastic
scattering: $E_0 = 0.145$\,MeV
\cite{ENSDF,NDS141}. The Brown-Rho parametrization in Eq.~(\ref{eq:BR})
is not able to reproduce the energy dependence of
the imaginary volume integrals and thus cannot be used for the extrapolation
of the potential to lower energies. This may -- at least partly -- explain the
problems with the prediction of the $^{144}$Sm\rag $^{148}$Gd cross section
\cite{Som98} using the potential derived from $^{144}$Sm\raa $^{144}$Sm
elastic scattering \cite{Mohr97} in combination with a Brown-Rho
parametrization of the imaginary volume integral. 

\begin{figure}[htb]
\includegraphics[width=\columnwidth,clip=]{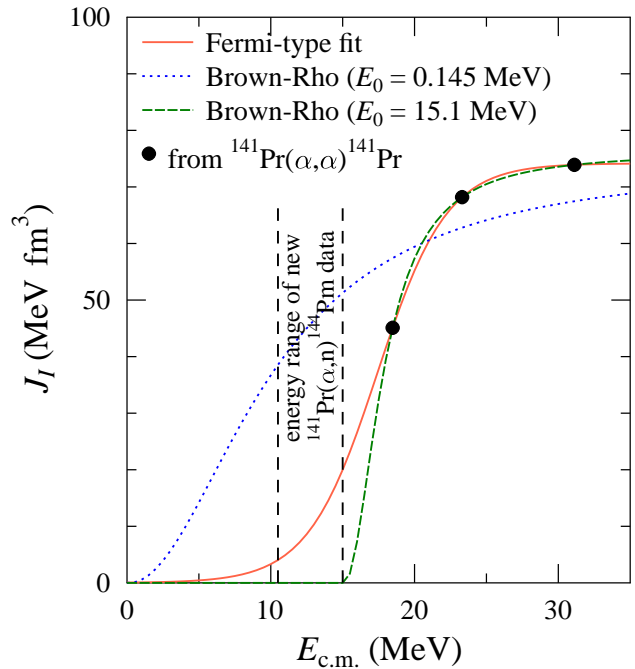}
\caption{
\label{fig:JI}
(Color online)
Energy dependence of the imaginary volume integral $J_I$ using the Fermi-type
parametrization of Eq.~(\ref{eq:fermi}) (full red line) and Brown-Rho
parametrizations of Eq.~(\ref{eq:BR}) with fixed $E_0 = 0.145$\,MeV (dotted
blue line) and $E_0 = 15.1$\,MeV (dashed green line)
with $E_0$ adjusted to the scattering data. The energy range of the recent
\pras \ran \pmiv\ experiment \cite{Sau11} is indicated by vertical
lines. Further discussion see text. 
}
\end{figure}

The failure of the Brown-Rho
parametrization is shown in Fig.~\ref{fig:JI}. Whereas the three parameters of
the Fermi-type function allow a perfect reproduction of the $J_I$ values from
elastic scattering, it is impossible to reproduce the steep rise of the $J_I$
data between about 15 and 25\,MeV with a Brown-Rho function and fixed $E_0 =
0.145$\,MeV. If the energy $E_0$ of the opening of the first non-elastic
channel is taken as an additional free parameter, the $J_I$ values from
elastic scattering 
can be reproduced with $J_I = 76.2$\,MeV\,fm$^3$, $E_0 = 15.1$\,MeV, and
$\Delta = 2.8$\,MeV. However, this result corresponds to a vanishing imaginary
part already in the energy range of the recent \pras \ran \pmiv\ experiment
\cite{Sau11} and thus predicts that the total reaction cross section
vanishes. It is clear that the Brown-Rho parametrization of the imaginary
volume integral $J_I$ cannot be used for the prediction of \al -induced
reaction cross sections of \pras .

\subsection{\al -decay of \pmv }
\label{sec:decay}
\al -decay provides a further opportunity to test \al -nucleus potentials at
low energies. For \pmv\ (as in most other cases) \al -decay is dominated by
the $\Delta L = 0$ transition from the \pmv\ ground state to the \pras\ ground
state; thus, analysis of \al -decay properties provides information on the
potential for the $L = 0$ partial wave (see also Sect.~\ref{sec:disc} for the
energy-dependent relevance of different partial waves for \stot ).
The mass range around $A \approx 150$ has been studied in
\cite{Mohr00} using the same type of folding potentials as in this
work. Although the  
\al -decay half-lives vary over many orders of magnitude between $^{148}$Gd,
$^{146}$Sm, and $^{144}$Nd (with their $N=82$-daughters $^{144}$Sm,
$^{142}$Nd, and $^{140}$Ce), it has been found in \cite{Mohr00} that the
preformation of the \al -particle in the decaying nucleus is between about 10
and 20\,\% within this model. 

\pmv\ with its $N=82$-daughter \pras\ has a weak \al -decay branch
of $(2.8 \pm 0.6) \times 10^{-9}$ \cite{ENSDF,NDS145,Nur62} which together with
the half-life of $T_{1/2} = 17.7 \pm 0.4$\,y \cite{ENSDF,NDS145,Bro59} translates
to a partial \al -decay half-life $T_{1/2}^\alpha = (2.0 \pm 0.4) \times
10^{17}$\,s. The 
\al -decay $Q$-value is $Q_\alpha = 2322.2 \pm 2.6$\,keV \cite{ENSDF}.
To repeat the \al -decay calculations of \cite{Mohr00} for \pmv ,
a real folding potential has been calculated at extremely low energies
(labelled ``\al\ decay'' in Table \ref{tab:pot}), and the \al -decay half-life
has been calculated using the semi-classical model of \cite{Gur87}. From the
ratio between the calculated half-life $T_{1/2}^{\alpha,{\rm{calc}}} = 2.49
\times 10^{16}$\,s and the experimental partial half-life a preformation of
$P = 12.5 \pm 2.7$\,\% is determined which is within the range of $10 -
20$\,\% for the neighboring \al\ emitters with $N=82$ daughters. Although
based on almost 50 years old data for the half-life \cite{Bro59} and the
\al\ branching \cite{Nur62}, this result nicely confirms the close
relationship between the various $N=82$ nuclei including $^{144}$Sm.

\section{\al -induced reactions on \pras }
\label{sec:reac}
As the optical potential for \pras -\al\ is completely fixed from the above
procedure in Sec.~\ref{sec:energy}, the calculation of the total reaction
cross section \stot\ is straightforward and does not require any further
parameter adjustment to experimental reaction data. First,
the obtained \stot $(E)$ is converted to the reduced cross section \sred\ and
compared to 
the \sred\ data for various projectiles and targets (full line in
Fig.~\ref{fig:sig_red}). It is obvious that the general behavior of the
\sred\ $vs.$\ $E_{\rm{red}}$ energy dependence is very nicely reproduced at
least down to $E_{\rm{red}} \approx 0.8$\,MeV corresponding to $E_{\rm{c.m.}}
\approx 14$\,MeV for the \pras -\al\ system under study in the work.

\begin{figure*}[htb]
\includegraphics[bbllx=35,bblly=20,bburx=785,bbury=425,width=0.8\textwidth,clip=]{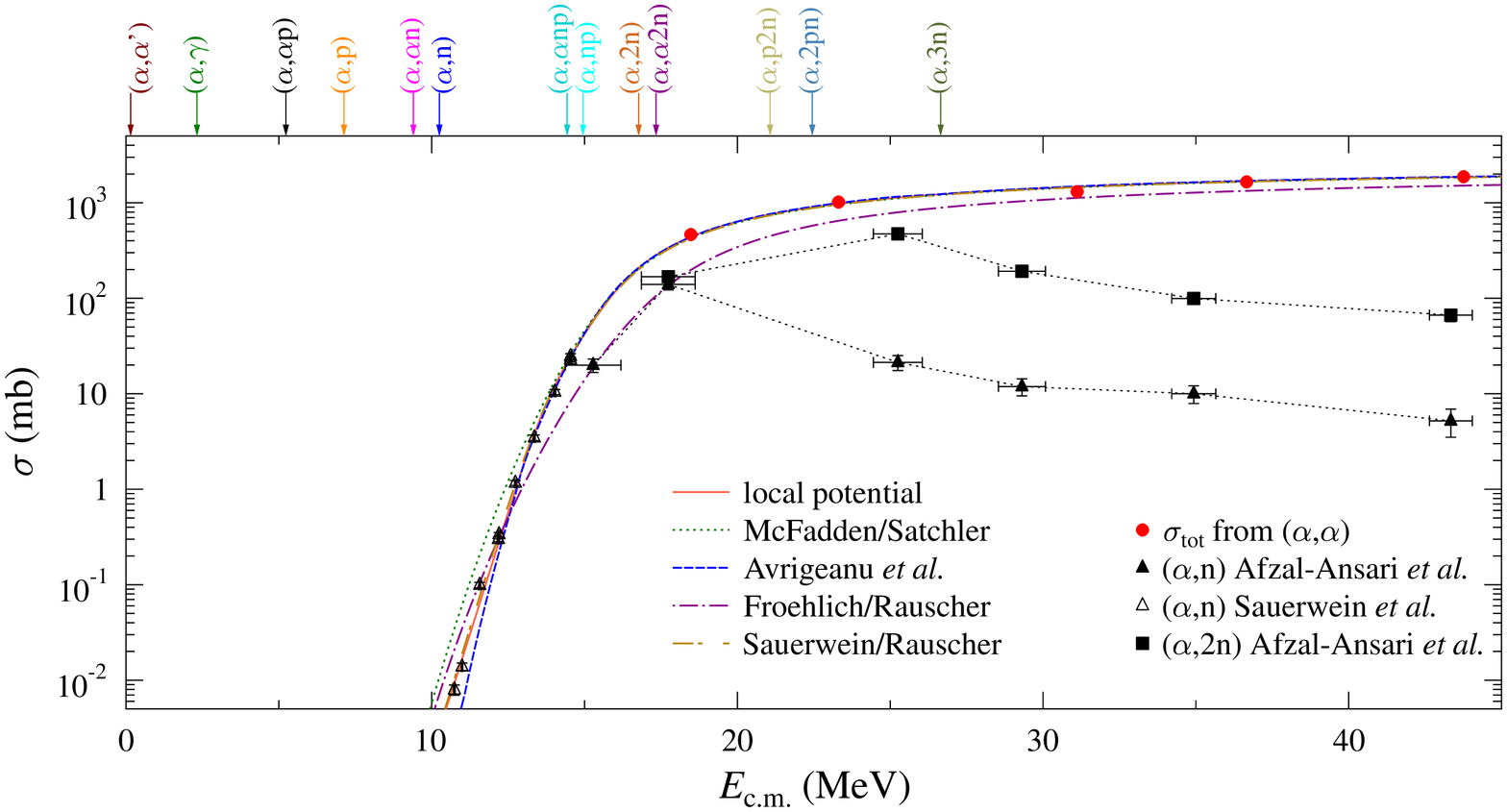}
\caption{
\label{fig:sigtot}
(Color online)
Total reaction cross sections \stot\ for \al -induced reactions on
\pras\ from elastic scattering (red dots) and cross sections of the \pras \ran
\pmiv\ (open and full triangles \cite{Afz05,Sau11}) and \pras \rann
\pmiii\ (full squares \cite{Afz05})
reactions. The data of \cite{Afz05} are connected by thin lines to guide the
eye. Calculations with different potentials
\cite{McF66,Avr10,Rau03,Sau11} agree well at higher energies (except
\cite{Rau03}), but disagree at lower energies (more details see also next
Fig.~\ref{fig:sfact}). The arrows on top indicate the thresholds for various
reaction channels.
}
\end{figure*}

\begin{figure}[htb]
\includegraphics[width=\columnwidth,clip=]{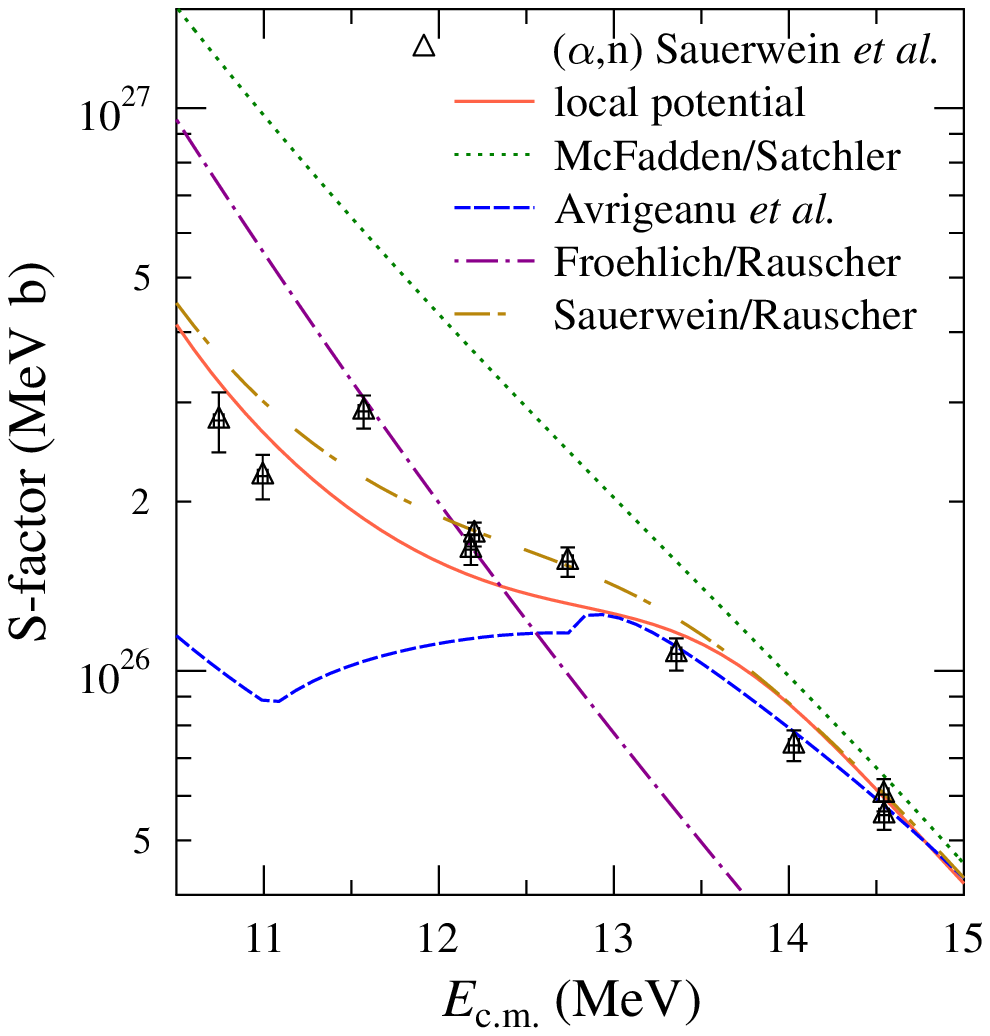}
\caption{
\label{fig:sfact}
(Color online)
Same as Fig.~\ref{fig:sigtot}, but shown as astrophysical \sfact\ in a narrow
energy window around the new experimental data of \cite{Sau11}. For
simplicity, the averaged \sfact\ at the mean energy (see Table III of
\cite{Sau11}) is shown here. The local potential (full red line) and the SAR
potential (dash-dotted golden) reproduce the data well, but the
global potentials fail to reproduce the energy dependence (MCF: dotted green;
AVR: dashed blue; FRR: dash-dotted dark-magenta).
}
\end{figure}

Next, the result for \stot\ is shown in
Fig.~\ref{fig:sigtot} in a wide energy range, and it is converted to the
astrophysical \sfact\ in Fig.~\ref{fig:sfact} for comparison with the new
\pras \ran \pmiv\ data of \cite{Sau11} for energies between 10 and 15\,MeV. 
The new experimental \sfact\ data decrease with energy by about a factor 5
from $S(E) \approx 3 \times 10^{26}$\,MeV\,b below 11\,MeV to $S(E) \approx
0.6 \times 10^{26}$\,MeV\,b at 14.5\,MeV. An
excellent reproduction of the new data of \cite{Sau11} is found.

\subsection{Total reaction cross section \sred\ and cross section of the \pras
  \ran \pmiv\ reaction}
\label{subsec:comparison}
The total reaction cross section \stot\ is given by the sum over all
non-elastic channels, i.e.\ it includes inelastic scattering, fusion, and all
transfer channels. This \stot\ has to be distributed among the open channels
(thresholds indicated in Fig.~\ref{fig:sigtot}) using e.g.\ the statistical
model. For the particular case of the \pras \ran \pmiv\ reaction in the energy
range of the recent experiment \cite{Sau11} as shown in Fig.~\ref{fig:sfact},
it is found that \stot\ is almost
identical to the \ran\ cross section because of the dominating neutron
emission channel (except very close above the \ran\ threshold at
10.2\,MeV). The proton emission from the compound nucleus \pmv\ is strongly
suppressed by the Coulomb barrier, and the 
ratio of \ran\ over \rag\ cross sections is large in this energy window. Using
the standard NON-SMOKER parameters \cite{Rau00} (\al -potential from
\cite{McF66}, nucleon 
potentials from \cite{Jeu77}, \g -ray strength function from \cite{McC81},
level density 
from \cite{Rau97}) the ratio is $\approx 4$ at the lowest energy of
\cite{Sau11} and exceeds 20 at 11.5\,MeV; i.e.\ neglecting the weak other open
channels like (\al ,\al n), the total reaction cross section \stot\ has to be
reduced by less than 5\,\% to obtain the \ran\ cross section above
11.5\,MeV and by $\approx 25$\,\% for the two lowest data points. The
standard TALYS \cite{tal07} calculation (\al -potential \cite{Wat58}, nucleon
potentials \cite{Kon03}, \g -ray strength function \cite{Kop90}, level density
\cite{Eri60,tal07}) predicts a ratio of $\approx 10$ between \ran\ and
\rag\ at the lowest energy of \cite{Sau11} and more than 100 at energies above
$\approx 12$\,MeV; i.e., the \ran\ cross 
section does not deviate by more than 1\,\% from \stot\ above 12\,MeV, and the
deviations at the lowest energies of \cite{Sau11} never exceed 10\,\%. Thus,
because only minor differences are predicted between the \ran\ cross section and
\stot\ in the energy range of \cite{Sau11} in all calculations
\cite{Rau00,tal07}, I restrict myself to the presentation of 
\stot\ in Figs.~\ref{fig:sigtot} and \ref{fig:sfact}. This avoids any
uncertainties from other sources (mainly from the neutron potential and the 
\g -ray strength function) which may be present at the lowest data points of
\cite{Sau11}. The minor differences between the \pras \ran \pmiv\ cross section
and \stot\ can be seen from a comparison of the calculations in
Fig.~\ref{fig:sfact} of this work and Figs.~8 and 9 of \cite{Sau11} (the same
scale has been chosen in all these figures for simple comparison).

From Fig.~\ref{fig:sigtot} it can be read that \stot\ is well reproduced from
all potentials under study at higher energies above 25 MeV (except the FRR
potential which underestimates \stot\ at all energies). Discrepancies become
visible at lower 
energies, see Fig.~\ref{fig:sfact}. The MCF potential shows an incorrect
energy dependence and strongly overestimates the data at lowest energies. The
AVR potential reproduces the experimental data in the higher energy range, but
underestimates the data 
at lower energies. The energy dependence calculated from the FRR potential is
also incorrect, and data at low (high) energies are over- (under-)
estimated. An energy-dependent potential was found in \cite{Sau11} which is
able to reproduce the new \pras \ran \pmiv\ data over the whole measured
energy range. The prediction of the new energy-dependent potential from the
present study which was adjusted only to elastic scattering data is very close
to the result of \cite{Sau11}, but does not require any adjustment to reaction
data.

It is interesting to note that the energy of the lowest data point in
Fig.~\ref{fig:sfact} corresponds to a reduced energy $E_{\rm{red}} \approx
0.62$\,MeV which is close to the lower end of Fig.~\ref{fig:sig_red}. The
reduced cross sections \sred\ at reduced energies below $E_{\rm{red}} =
1$\,MeV are thus almost identical for \pras\ + \al\ and $^{16}$O + $^{138}$Ba;
in both cases the projectile is a doubly-magic nucleus and the target is
semi-magic with $N=82$. 

\subsection{Sensitivity of the total reaction cross section \sred\ to
  variations of the parameters of the local potential}
\label{sec:sensitivity}
\begin{figure}[htb]
\includegraphics[width=\columnwidth,clip=]{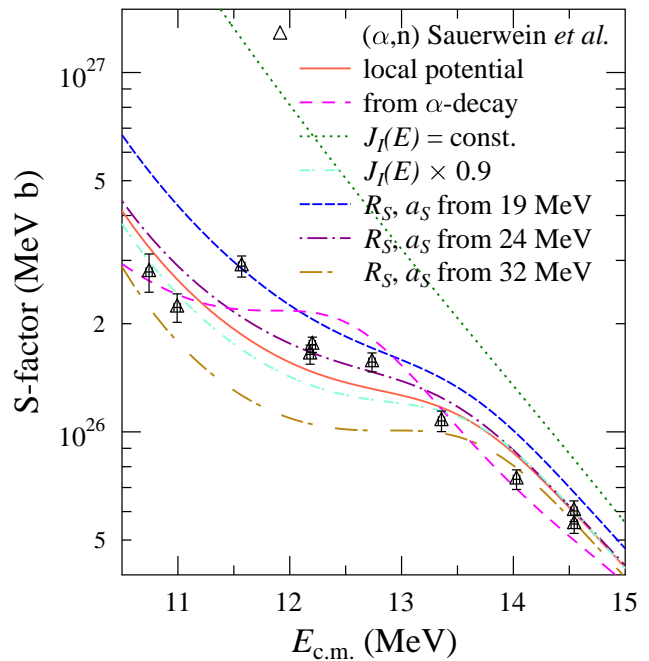}
\caption{
\label{fig:sens}
(Color online)
Sensitivity of the astrophysical \sfact\ to variations of the potential
parameters. The full red line is identical to the previous
Fig.~\ref{fig:sfact}. A reduced real potential (derived from \al -decay, see
Sect.~\ref{sec:decay}) shows a slighly different energy dependence (dashed
magenta). Neglecting the energy dependence of the imaginary part leads to
strong overestimation of the experimental cross sections (green dotted),
whereas small modifications of the imaginary part (10\,\% reduction of the
strength or variation of geometry parameters) do not affect the calculated
\stot\ strongly. Further discussion see text.
}
\end{figure}

It is difficult to provide a precise error bar for the calculated total
reaction cross section \stot\ in Figs.~\ref{fig:sigtot} and \ref{fig:sfact},
but an estimate of the uncertainties can be given by reasonable variations of
the parameters of the local potential which are the geometry (in particular
of the imaginary part) and the energy dependence. The results of the following
sensitivity study are shown in Fig.~\ref{fig:sens}. The local potential
calculation (full red line, identical to Fig.~\ref{fig:sfact}) is the
reference for the sensitivity study.

The real part of the potential is nicely constrained from the folding
calculation. In addition to the scattering data, further information on the
real part of the potential at very low energies can be extracted from the
analysis of the \al -decay of \pmv\ (see Sect.~\ref{sec:decay}).
The parameters of the potential for the \al -decay calculations remain close
to the real part of the scattering potential. The volume integral $J_R$ is
about 10\,\% lower than the value found for the 19\,MeV scattering data. If
this lower real part of the potential is used for the calculations of
\stot\ instead of the 19\,MeV real part (as derived in Sec.~\ref{sec:energy}),
\stot\ shows a slightly different energy dependence but
does not change by more than about 40\,\% in the energy range under
study (dashed fuchsia line in Fig.~\ref{fig:sens}). 
Only at very low energies below 10\,MeV (i.e., below the shown energy range
of Fig.~\ref{fig:sens}), the lower real potential with
its resulting higher effective Coulomb barrier results in a lower \stot . But
in any case the deviations remain below 50\,\%, thus confirming that the real
part of the potential is relatively well-defined and does not lead to
big uncertainties in the calculation of \stot\ in the energy range under
analysis. 

It should be noted that the \al -decay potential is adjusted at the
decay energy of 2.3\,MeV far below the energy range under study; thus, using
the \al -decay potential for the calculation of \stot\ is an extremely careful
estimate for the uncertainty of \stot\ on a variation of the real
potential. I do not show calculations with increased strength of the real
part because volume integrals significantly above 350\,MeV\,fm$^3$ have not
been observed in \al -scattering of semi-magic nuclei
\cite{Atz96,Ful01,Gal05,Kis09}. This finding is supported theoretically
by the fact that
dispersion relations lead to a reduction of real potential at very low
energies (see e.g.\ Fig.~11 of \cite{Atz96} or Fig.~12 of \cite{Abe93}).

The influence of the imaginary part on the calculated total reaction cross
section \stot\ is significant. In particular, a reasonable energy dependence
is essential for the reproduction of the experimental data. If the energy
dependence of the imaginary part is ignored and the saturation value $J_{I,0}$
is used instead of the energy dependence of Eq.~(\ref{eq:fermi}), then
\stot\ is overestimated with increasing discrepancy to the experimental data
at lower energies (green dotted line in Fig.~\ref{fig:sens}). A similar
behavior has been found for the energy-independent MCF potential. However,
small variations of the imaginary strength result only in minor modifications
of \stot : a reduction of $J_I(E)$ by 10\,\% leads to slightly reduced
\stot\ with very similar energy dependence as the reference calculation
(dash-dotted aquamarine).

Next the geometry parameters of the surface Woods-Saxon potential in the
imaginary part were taken from the three angular distributions at 19, 24, and
32\,MeV (see Table \ref{tab:pot}) instead of their average values $R_S =
1.353$\,fm and $a_S = 0.530$\,fm. The depth $W_S$ has been adjusted to the
same volume integral in Eq.~(\ref{eq:fermi}) as in the reference
calculation. The geometry from the 19\,MeV angular distribution leads to
somewhat larger \stot\ which is a consequence of the larger $a_S =
0.581$\,fm (short-dashed blue). The result from the 24\,MeV geometry is almost
identical to the reference calculation (dash-dotted magenta), and the result
from the 32\,MeV geometry is somwehat smaller than the reference (dash-dotted
golden). The uncertainty of the geometry of the imaginary potential leads to
uncertainties of \stot\ of the order of 30\,\%. It has to be pointed out that
such a small uncertainty can only be achieved because the geometry parameters
are well-defined from several angular distributions at energies around and
slightly above the Coulomb barrier.

Summarizing the above sensitivity study, it is found that the uncertainty of
\stot\ from the present local potential is much smaller than the variations
between the different predictions from global potentials. The influence of the
geometry of the imaginary part is not dramatic as long as the geometry is
well-defined from low-energy scattering data. However, the energy dependence
of the strength $J_I$ of the imaginary part is an essential ingredient for the
prediction of reaction cross sections below the Coulomb barrier.

\section{Discussion}
\label{sec:disc}
Let me start the discussion with a few general remarks on statistical model
calculations. In the statistical model the cross section for a reaction is
given by the product of the compound formation cross section and the decay
branching of the decaying compound nucleus into the particular channel. The
full formalism can be found e.g.\ in \cite{Rau00,Rau11}. The formation cross
section is calculated from transmission coefficients using the optical
potentials, i.e.\ it is the total reaction cross section \stot\ in
Eq.~(\ref{eq:stot}). The decay branching is also calculated from transmission
coefficients for particle channels and from the $\gamma$-ray strength function
for the photon (capture) channel. It is obvious that the correct reproduction
of the total reaction cross section \stot\ is the basic prerequisite for a
successful prediction of any reaction cross section in the statistical model
and should thus always be discussed first in presentations of statistical
model calculations. Unfortunately, this is not always the case.

The total reaction cross section \stot\ is then distributed among all open
channels. It is important to study the influence of all open channels and
their relevance for the decay branching of the compound nucleus. This is
nicely illustrated in Fig.~\ref{fig:sigtot}. Close above the opening of the
\rann\ channel around 17\,MeV, the \rann\ channel becomes stronger than the
\ran\ channel which was dominant from close above its threshold at 10.2\,MeV
up to 17\,MeV. If an open channel, e.g.\ the \pras \rann \pmiii\ channel above
17\,MeV, is not taken into account in a statistical model calculation, then
all the calculated \pras \raX\ cross sections above 17\,MeV must be
overestimated. However, such a shortcoming of neglected open channels may be
partly compensated by using special potentials for the particular
\raX\ reaction under study. 

After these general remarks let me be more specific for \al -induced reactions
on \pras . The total reaction cross section \stot\ is well understood and has
relatively small uncertainties for energies above the Coulomb barrier. This
can be seen from the comparison of predicted \stot\ from the various global
potentials \cite{McF66,Avr10,Rau03} to the experimental result from the
analysis of the elastic scattering angular distributions (see Table
\ref{tab:sigtot}). All potentials (except the FRR potential) reproduce the
experimental \stot\ within minor uncertainties of typically a few per cent.
\begin{table}[tbh]
\caption{\label{tab:sigtot}
Experimental total reaction cross sections \stot\ (in mb) derived from \pras
\raa \pras\ elastic scattering angular distributions (from Table
\ref{tab:pot}), compared to the phase shift analysis 
(``PSA''), predictions from the global \al -nucleus potentials MCF, AVR, and
FRR \cite{McF66,Avr10,Rau03}, and the energy-dependent potential SAR of
\cite{Sau11} which was adjusted to reproduce the \pras \ran \pmiv\ cross
section at low energies.
}
\begin{center}
\begin{tabular}{cr@{$\pm$}lrrrrr}
\multicolumn{1}{c}{$E_{\rm{c.m.}}$ (MeV)}
& \multicolumn{2}{c}{exp.}
& \multicolumn{1}{c}{PSA}
& \multicolumn{1}{c}{MCF}
& \multicolumn{1}{c}{AVR}
& \multicolumn{1}{c}{FRR}
& \multicolumn{1}{c}{SAR} \\
\hline
18.48  &  465  & 31  &  433  &  425  &  437  &  197  &  424 \\
23.31  & 1018  & 31  & 1003  &  957  &  993  &  648  &  957 \\
31.12  & 1306  & 39  & 1310  & 1452  & 1487  & 1125  & 1452 \\
36.66  & 1654  & 50  & 1654  & 1662  & 1696  & 1333  & 1662 \\
43.76  & 1874  & 56  & 1873  & 1841  & 1872  & 1514  & 1841 \\
\hline
\end{tabular}
\end{center}
\end{table}

\begin{figure}[htb]
\includegraphics[width=\columnwidth,clip=]{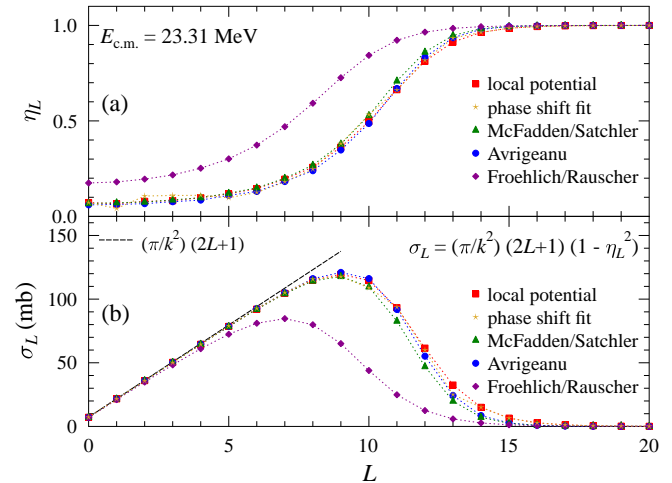}
\caption{
\label{fig:eta_reac-e24}
(Color online)
Reflexion coefficients $\eta_L$ (upper, (a)) and contribution $\sigma_L$ of the
$L$-th partial wave to the total reaction cross section \stot\ (lower, (b);
linear scale). The thin dotted lines connect the data points to guide the
eye. Further discussion see text.
}
\end{figure}

The excellent reproduction of \stot\ for most global potentials at energies
above the Coulomb barrier can 
be explained by a look at the reflexion coefficients $\eta_L$ and the
contribution $\sigma_L$ of the $L$-th partial wave to \stot\ (see
Fig.~\ref{fig:eta_reac-e24} for the example of the $E_{\rm{c.m.}} =
23.31$\,MeV data). For small angular momenta $L$ the reflexion 
coefficients are close to $\eta_L \approx 0$; thus, $\sigma_L$
increases linearly proportional to $2L+1$ (dashed line in
Fig.~\ref{fig:eta_reac-e24}). All 
global potentials with a reasonable real part and a sufficiently strong
imaginary part, i.e.\ the MCF, 
AVR, and SAR potentials, predict the same $\sigma_L$ for small
$L$. For large $L$ (corresponding to large impact parameters) the $\eta_L$
approach unity, and the $\sigma_L$ vanish (i.e., partial waves with large
angular momenta are practically not absorbed by any imaginary potential with a
limited radial range). Differences in $\sigma_L$ appear only for a few partial
waves between $L \approx 8$ and $L \approx 14$ (for the $E_{\rm{c.m.}} =
23.31$\,MeV case) where the different potentials predict different 
$\eta_L$ and thus different $\sigma_L$. It becomes obvious that relatively
small differences in $\eta_L$ for a very limited number of partial waves
around $L \approx 10$ cannot lead to major differences for the total reaction
cross section \stot\ which is simply given by the sum over all $\sigma_L$. The
good reproduction of the total reaction cross section \stot\ from different
global potentials is thus simply a consequence of intrinsic properties of the
scattering matrix as long as the energy exceeds the Coulomb barrier.

\begin{figure}[htb]
\includegraphics[width=\columnwidth,clip=]{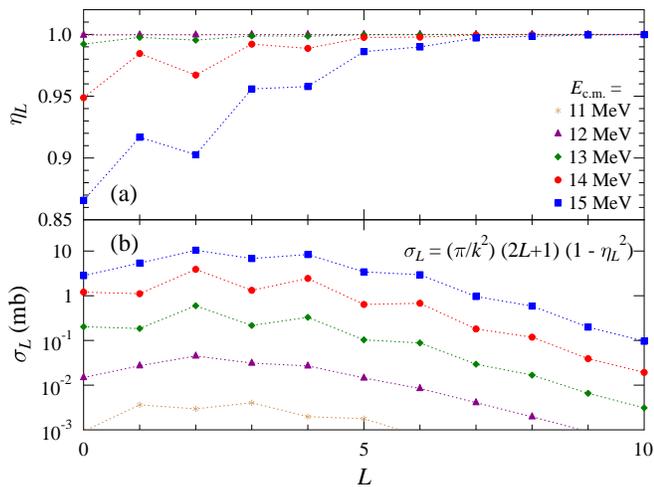}
\caption{
\label{fig:eta_reac-low}
(Color online)
Reflexion coefficients $\eta_L$ (upper, (a)) and contribution $\sigma_L$ of the
$L$-th partial wave to the total reaction cross section \stot\ (lower, (b);
logarithmic scale) for low energies between 11 and 15\,MeV. Note the different
scale for $\eta_L$ in Figs.~\ref{fig:eta_reac-e24} and
\ref{fig:eta_reac-low}. The thin dotted lines connect the data points to guide 
the eye. The $\eta_L$ at 11 and 12\,MeV are very close to unity for all $L$;
thus, the 11\,MeV data disappear behind the 12\,MeV data in the upper
diagram. Differences between 
the 11 and 12\,MeV data become visible only in the presentation of $\sigma_L$
which are proportional to $(1-\eta_L^2)$.
The $2L+1$ proportionality of $\sigma_L$ for small $L$ (as seen at higher
energies in Fig.~\ref{fig:eta_reac-e24}) vanishes completely at energies
below the Coulomb barrier. Further discussion see text.
}
\end{figure}

The situation changes dramatically at lower energies far below the Coulomb
barrier. Here only the lowest angular momenta (below $L \approx 5$) are
affected, show $\eta_L < 1$, and thus contribute to $\sigma_L$; but even these
$\eta_L$ remain much larger than zero and do even approach $\eta_L \approx 1$
for very low energies (see Fig.~\ref{fig:eta_reac-low} for energies
$11\,{\rm{MeV}} \le E_{\rm{c.m.}} \le 15\,{\rm{MeV}}$, i.e.\ the energy range
of the new \pras \ran \pmiv\ data of \cite{Sau11}). Here the trivial $2L+1$
proportionality of the $\sigma_L$ vanishes. Now the relevant $\sigma_L$ depend
sensitively on details of the potential. In particular, the strength of the
imaginary part for large radii has strong impact on the resulting cross
sections $\sigma_L$. Under these conditions it becomes mandatory that the
imaginary potential has the proper geometry (e.g.\ fixed from scattering
data at slightly higher energies) and the proper strength (using a realistic
energy dependence, e.g.\ from Eq.~(\ref{eq:fermi}), which can be adjusted to
scattering data and/or properly chosen reaction data). As pointed out in
Sec.~\ref{sec:scat}, the real part of the potential has only minor
uncertainties in shape and strength because it is calculated from a folding
procedure; furthermore, the analysis of \al -decay properties provides a test
for the $L=0$ potential (see Sect.~\ref{sec:decay}). 

The above discussion explains why all global potentials are able to reproduce
total reaction cross sections \stot\ above the Coulomb barrier, but may fail
to reproduce \stot\ below the Coulomb barrier. Thus, it is not surprising that
the global potentials show a significant scatter in the predictions of the
new experimental \pras \ran \pmiv\ data of \cite{Sau11} (see
Fig.~\ref{fig:sfact}). 

For the energy-independent MCF potential the reason for the discrepancy is
obvious. The missing energy dependence of the MCF potential leads to an
overestimation of the imaginary part of the potential at low energies and thus
to an overestimation of the \pras \ran \pmiv\ cross section. As expected, the
discrepancy between the MCF prediction and the experimental data increases
with decreasing energy. A similar behavior has been found in other cases, see
e.g.\ the $^{144}$Sm\rag $^{148}$Gd reaction \cite{Som98}, the $^{112}$Sn\rag 
$^{116}$Te reaction \cite{Ozk07}, or the $^{106}$Cd\rag $^{110}$Sn reaction
\cite{Gyu06}. 

It is difficult to make such general statements on the many-parameter AVR
potential. This potential has been adjusted to a huge data base of elastic
scattering and reaction data \cite{Avr09}, and excellent agreement has been
found for many reactions especially in the $A \approx 100$ range. The AVR
potential shows the best agreement of the global potentials with the elastic
scattering angular distributions, and it reproduces the \pras \ran \pmiv\ data
well at higher energies above 13\,MeV. However, it underestimates the data at
lower energies significantly.

The FRR potential has been optimized for reaction data at low energies. Its
parameters are energy-independent and close to the MCF potential in most
cases. Similar to the MCF case, because of its energy independence it cannot
be expected that the FRR
potential is able to reproduce simultaneously elastic scattering data above
the Coulomb barrier and reaction data below the Coulomb barrier. Despite the
big success in predicting reaction cross sections at low energies, the simple
FRR potential cannot be the basis for a global potential in a broad energy
range.

The MCF potential with an additional energy dependence of the depth of the
imaginary potential has been used in \cite{Sau11} to reproduce their new \pras
\ran \pmiv\ data. Although this is in principle the best way to improve the
MCF potential, there are two disadvantages of the SAR potential. Obviously,
reaction data are required to fit the energy dependence of the SAR
potential. And in addition, the chosen underlying MCF potential uses a volume
Woods-Saxon imaginary part which has been found -- according to recent studies
\cite{Gal05,Kis11a,Guo11} -- to be probably not fully adequate for low
energies below the 
Coulomb barrier; here surface Woods-Saxon potentials should be
preferred. Consequently, the SAR potential is not able to reproduce the
elastic scattering angular distributions. It turns out that the angular
distributions of the SAR and MCF potentials are almost identical above
25\,MeV because the MCF and SAR potentials are almost identical there. Even
at 19\,MeV where the imaginary depth of the SAR potential is about 25\,\%
lower than the MCF depth, 
the calculated angular distributions of MCF and SAR agree within a few per
cent, i.e.\ within a linewidth in Fig.~\ref{fig:scat}; therefore, the SAR
potential is not shown in Fig.~\ref{fig:scat}.

The energy-dependent potential of this work has been derived from elastic
scattering data. Local optical potentials have been derived from the available
angular distributions between 19 and 45\,MeV by fitting the parameters of the
real part (strength and width of the folding potential) and the imaginary part
(Woods-Saxon parametrization). Because of the small variation of the found
parameters at the different energies (see Table \ref{tab:pot}), these
parameters could be combined to derive a common potential with
an energy-dependent depth of the imaginary surface potential. This common
potential maintains the good reproduction of the elastic scattering data and
the derived total reaction cross sections \stot . In
addition, it turns out that this potential is able to predict the cross
section of the \pras \ran \pmiv\ cross section at lower energies, in
particular in the energy range of the new experimental data of \cite{Sau11}
between 10 and 15\,MeV (see Fig.~\ref{fig:sfact}). Thus, the new potential is
able to describe \pras \raa \pras\ elastic scattering data at energies around
and above the Coulomb barrier simultaneously with \pras \ran \pmiv\ reaction
data below the Coulomb barrier.

The parametrization of this new potential, i.e.\ a double folding potential in
the real part and a surface Woods-Saxon potential with fixed geometry and
energy-dependent depth, should be tested in further cases in forthcoming
work. For a detailed study of a particular target nucleus, several elastic
scattering angular distributions are required to fix the shape and the energy
dependence of the imaginary potential. However, the found geometry parameters
in the present \pras -\al\ case ($R_S = 1.353$\,fm and $a_S = 0.530$\,fm) are
close 
to standard values; it should be possible to apply these parameters to other
target nuclei. The main problem will be the determination of the energy
dependence of the depth of the imaginary surface potential. Here
considerations similar 
to \cite{Avr09,Avr10,Sau11} will be helpful where parameters of the imaginary
part have been put in relation to the Coulomb barrier. It is obvious that the
predictions from such a potential have to be compared to experimental reaction
data below the Coulomb barrier which are -- despite significant improvement in
the recent years -- still not widely available.

The previous discussion may also be summarized to provide a general recipe to
find a widely useful potential. First of all, the potential has to reproduce
the total reaction cross section \stot\ at energies above the Coulomb
barrier. However, because of the $2L+1$ proportionality of $\sigma_L$ for
small $L$, this requirement is almost trivial and can be fulfilled by many
potentials. Next, among 
the many \al -nucleus potentials which are able to fulfill this first
condition, these potentials have to be selected which are in addition able to
reproduce the elastic scattering angular distributions. This ensures the
correct strength and geometry of the chosen potential (in particular of the
imaginary part) which is essential for the prediction of \stot\ below the
Coulomb barrier. This conclusion is also a strong motivation to extend the
measurements of elastic scattering angular distributions.

\section{Summary and Conclusions}
\label{sec:summ}
A new local potential for the system \pras -\al\ has been derived from
elastic scattering angular distributions. The derived potential is able to
predict the total reaction cross section \stot\ which is dominated by the
\pras \ran \pmiv\ cross section in the energy range of the new experimental
data of \cite{Sau11}. Thus, the new potential for \pras -\al\ is able for
the first time to describe simultaneously elastic scattering data around and
above the Coulomb barrier and reaction data below the Coulomb barrier. Such a
simultaneous description was not achieved and/or aimed for in earlier work
using the global potentials of \cite{McF66,Avr10,Rau03} or the local potential
suggested in \cite{Sau11}.

Reasons for the partial success at higher energies and failure at lower
energies of global potentials are carefully analyzed by studying the reflexion
coefficients $\eta_L$ and the contribution $\sigma_L$ 
of the $L$-th partial wave to the total reaction
cross section \stot . It is found that above the Coulomb barrier $\sigma_L$ for
small angular momenta are predicted correctly by most global potentials
because of a simple $2L+1$ proportionality which is almost independent of
details of the optical potential. This results in minor differences for the
total reaction cross section \stot\ which have to arise from a few partial
waves around $L \approx 10$ for the \pras-\al\ case under study in this
work. However, at low energies below the Coulomb barrier this simple $2L+1$
proportionality of $\sigma_L$ for small $L$ vanishes, the $\sigma_L$ become
sensitive to 
details of the potential, and consequently predictions for \stot\ from global
potentials show a huge spread. Precise scattering data at different energies
around and close above the Coulomb barrier are required to determine a
potential (in particular the shape and strength of its imaginary part) which
is able to predict total reaction cross sections \stot\ not only 
above, but also below the Coulomb barrier.

Because the new local potential has a well-defined geometry derived from
elastic scattering, its extrapolation to low energies and its predictions for
the $\sigma_L$ and the resulting \stot\ should be more reliable than
predictions from other global potentials. The parametrization of the new local
potential may finally become the basis for a new global potential to solve or
at least reduce the long-standing problem of \al -nucleus potentials at low
energies below the Coulomb barrier and the resulting uncertainties for the
prediction of \al -induced reaction cross sections at astrophysically relevant
energies.

\begin{acknowledgments}
I thank P.\ Guazzoni and L.\ Zetta for providing their experimental \pras \raa
\pras\ scattering data before publication and R.\ Lichtenth\"aler for
providing the phase shift fitting code ``PARA'' of \cite{Chi96}. Encouraging
discussions with A.\ Sauerwein, A.\ Zilges, Zs.\ F\"ul\"op, and T.\ Rauscher
are gratefully acknowledged. I thank N.\ {\"O}zkan and T.\ G{\"u}ray for the
invitation to the Istanbul workshop on \ppro\ nucleosynthesis which was the
basis for this study. This work was supported by OTKA (NN83261).
\end{acknowledgments}

\end{document}